\documentclass[a4paper, 10pt, conference]{ieeeconf}      

\IEEEoverridecommandlockouts                              

\usepackage{amsmath} 
\usepackage{amssymb}  
\usepackage{bm} 
\usepackage{nicefrac}
\usepackage[nohyperlinks, printonlyused, withpage, nolist]{acronym} 
\usepackage{nameref} 
\usepackage{multirow} 
\usepackage{booktabs} 
\usepackage{array} 
\usepackage{paralist} 
\usepackage{hyperref} 
\usepackage{setspace} 
\usepackage{pdfpages} 

\usepackage{tikz}
\usepackage{pgfplots}
\pgfplotsset{compat=newest}
\usetikzlibrary{plotmarks}
\usetikzlibrary{arrows.meta}
\usetikzlibrary{shapes.geometric,shapes.symbols,fit,positioning,shadows,calc}
\usepgfplotslibrary{patchplots}
\usepackage{grffile}
\usepackage{xcolor}

\title{\LARGE \bf
Environment Modeling Based on Generic Infrastructure Sensor Interfaces Using a Centralized Labeled-Multi-Bernoulli Filter*
}

\author{Martin Herrmann$^{1}$, Johannes M{\"u}ller$^{1}$, Jan Strohbeck$^{1}$ and Michael Buchholz$^{1}$
\thanks{*Part of this work was financially supported by the Federal Ministry of Economic Affairs and Energy of Germany within the program "Highly and Fully Automated Driving in Demanding Driving Situations" (project MEC-View, grant number 19A16010I).}
\thanks{*Part of this work has been conducted as part of ICT4CART project which has received funding from the European Union's Horizon 2020 research \& innovation programme under grant agreement No. 768953. Content reflects only the authors' view and European Commission is not responsible for any use that may be made of the information it contains.
}
\thanks{$^{1}$The authors are with Institute of Measurement, Control, and Microtechnology, Ulm University, Germany
{\tt\footnotesize {\{firstname.lastname\}}@uni-ulm.de}}%
}

\newcommand\copyrighttext{%
    \footnotesize Copyright $\copyright$ 2019 IEEE.
    Personal use of this material is permitted.
    Permission from IEEE must be obtained for all other uses, in any current or future media, including reprinting/republishing this material for advertising or promotional purposes, creating new collective works, for resale or redistribution to servers or lists, or reuse of any copyrighted component of this work in other works.}%
\newcommand\copyrightnotice{%
    \begin{tikzpicture}[remember picture,overlay]%
    \node[anchor=south,yshift=10pt] at (current page.south) {\fbox{\parbox{\dimexpr\textwidth-2cm}{\copyrighttext}}};%
    \end{tikzpicture}%
    \vspace{-10pt}%
}

\begin{document}

\maketitle
\copyrightnotice
\thispagestyle{empty}
\pagestyle{empty}


\begin{acronym}
\acro{LMB}[LMB]{Labeled Multi-Bernoulli}
\acro{mMTC}[mMTC]{massive Machine Type Communications}
\acro{URLLC}[URLLC]{Ultra-Reliable and Low Latency Communications}
\acro{V2V}[V2V]{Vehicle-to-Vehicle}
\acro{V2X}[V2X]{Vehicle-to-Anything}
\acro{MEC}[MEC]{Multi-access Edge Computing}
\acro{ETSI}[ETSI]{European Telecommunications Standards Institute}
\acro{UE}[UE]{User Equipment}
\acro{eNB}[eNB]{evolved Node B}
\acro{5G}[5G]{5th Generation}
\acro{CAM}[CAM]{Cooperative Awareness Message}
\acro{CPM}[CPM]{Collective Perception Message}
\acro{FOV}[FOV]{Field of View}
\acro{FISST}[FISST]{Finite Set Statistics}
\acro{RFS}[RFS]{Random Finite Set}
\acro{JIPDA}[JIPDA]{Joint Integrated Probabilistic Data Association}
\acro{MHT}[MHT]{Multi-Hypothesis Tracking}
\acro{CPHD}[CPHD]{Cardinalized Probability Hypothesis Density}
\acro{GLMB}[GLMB]{Generalized Labeled Multi-Bernoulli}
\acro{UKF}[UKF]{Unscented Kalman Filter}
\acro{CTRV}[CTRV]{Constant Turn Rate and Velocity}
\acro{CTRA}[CTRA]{Constant Turn Rate and Acceleration}
\acro{GPS}[GPS]{Global Positioning System}
\acro{LTE}[LTE]{Long Term Evolution}
\acro{LTE-A}[LTE-A]{LTE-Advanced}
\acro{CI}[CI]{Covariance Intersection}
\acro{GCI}[GCI]{Generalized Covariance Intersection}
\acro{ITS}[ITS]{Intelligent Transportation System}
\acro{G5}[ITS-G5]{ITS-G5}
\acro{OSPAT}[OSPAT]{Optimal Sub-pattern Assignment metric for track}
\acro{MSE}[MSE]{Mean Squared Error}
\acro{iid}[i.i.d.]{independent and identically distributed}
\end{acronym}


\begin{abstract}
Urban intersections put high demands on fully automated vehicles, in particular, if occlusion occurs. In order to resolve such and support vehicles in unclear situations, a popular approach is the utilization of additional information from infrastructure-based sensing systems. However, a widespread use of such systems is circumvented by their complexity and thus, high costs. Within this paper, a generic interface is proposed, which enables a huge variety of sensors to be connected. The sensors are only required to measure very few features of the objects, if multiple distributed sensors with different viewing directions are available. Furthermore, a \ac{LMB} filter is presented, which can not only handle such measurements, but also infers missing object information about the objects' extents. The approach is evaluated on simulations and demonstrated on a real-world infrastructure setup.
\end{abstract}


\section{INTRODUCTION} \label{introduction}
Suppose a fully automated vehicle heading towards an urban, occluded and busy T-junction or intersection without having right of way. The vehicle's goal is to merge into the major road in a maximally safe and economic manner, hence it must be aware of the static and dynamic environment. However, this may be infeasible due to situation-dependent adverse mounting positions of the vehicle's sensors, or occlusion by static environment, such as buildings, signs or plants, or dynamic objects, e.g. other vehicles. This weakness is stressed out even more in complex urban environments, which put even higher requirements on intelligent vehicles. To overcome the disadvantages of such ego-centered perception, infrastructure-based cooperative surveillance systems can be utilized, which benefit from exposed sensing positions and knowledge of the local area.

\begin{figure}[tbp]
\centerline{
\includegraphics[width=0.8\linewidth]{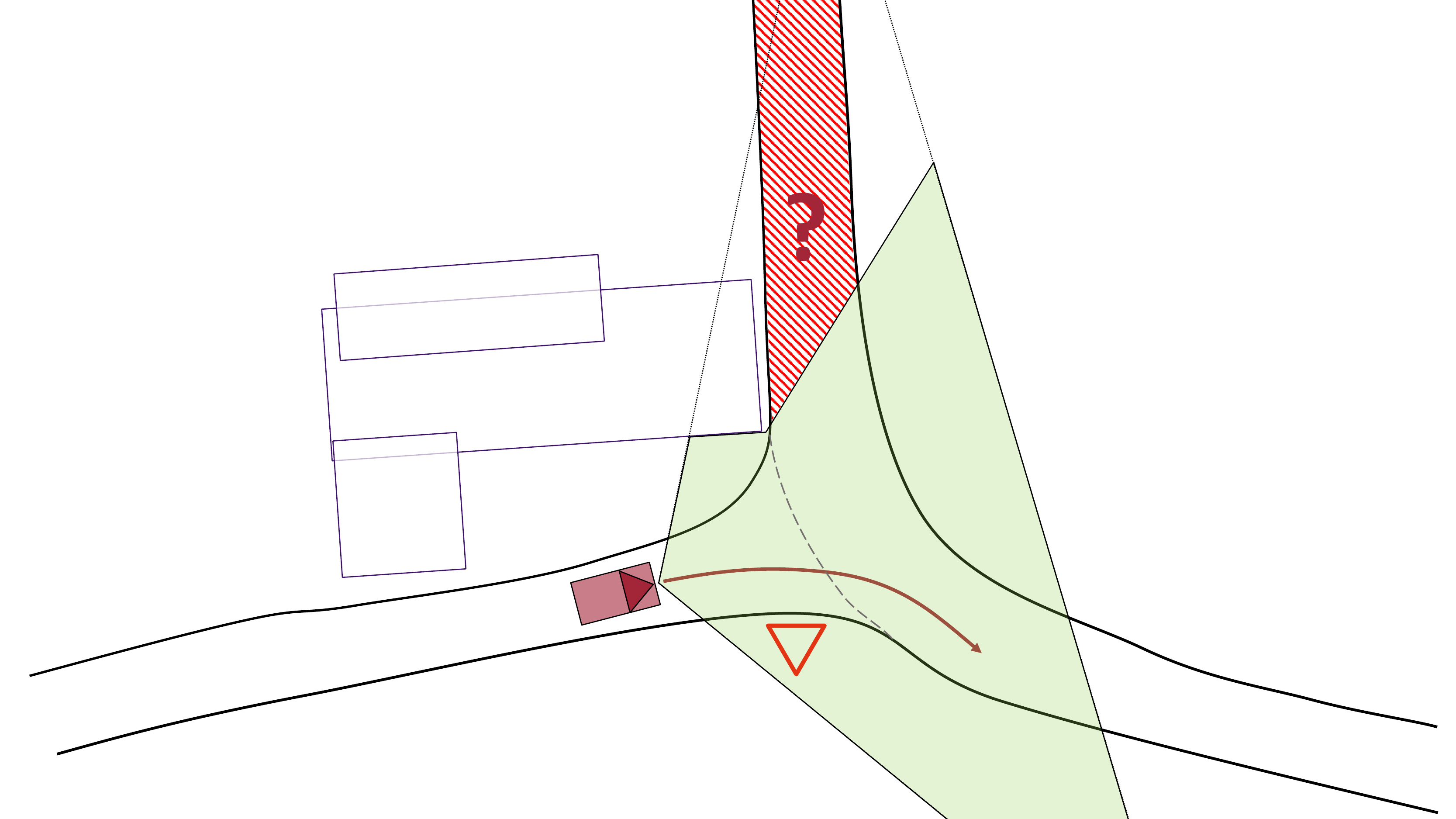}}
\caption{T-junction in Ulm-Lehr with a fully automated vehicle without right of way (red vehicle) and heavy occlusion due to a building.}
\label{fig:intersection_model}
\end{figure}

Recent advances in mobile radio technology cleared the way for a widespread use of such systems, based on distributed but radio-link connected sensors (and vehicles) in the public transportation system. A number of other projects, such as Ko-PER \cite{Co-PER}, DRIVE C2X \cite{drive_C2X} or simTD \cite{simTD} have addressed similar problems assuming different levels of automation, however, none of them explored a centralized \ac{MEC}-based approach. The idea of a \ac{MEC} server \cite{ETSI_MEC}, which is located close to the site, is part of the \ac{5G} radio network standard and allows for communication with ultra low latency. This puts the basis of the projects MEC-View \cite{mecview} and ICT4CART \cite{ICT4CART}. Within both projects, one aim is to enable connected and intelligent vehicles to perform advanced and economic maneuvers, such as accurately targeting a gap between vehicles with right of way in advance, without the need of a full stop at the stop line \cite{digital_mirror}. The problem is made even more complicated, since an unsignalized T-junction is considered, where vehicles without right of way suffer from heavy environment occlusion due to a large building, as shown in Fig.~\ref{fig:intersection_model}. A busy street in Ulm-Lehr was chosen as real test site to demonstrate the mentioned use-case. The chosen T-junction qualifies perfectly due to the occlusion by close buildings. It is equipped with several monocular cameras and lidars, which are connected to the \ac{MEC} server to surveil the intersection.

Within the scope of this work, the problem of environment modeling is tackled under the ambition of developing a centralized, infrastructure-based system, which follows the guidelines of being cheap and easy to install, adaptive to all kinds of sensors and arbitrarily structured urban environments, scalable, and reliable. Moreover, due to the generality against the sensor type, the problem of infering information about objects from incomplete measurements is covered. Therefore, resulting requirements on the system are:
\begin{enumerate}[i)]
\item Various different off-the-shelf sensors have to be usable, to keep installation and operating costs low. Since such sensors may show very limited measurement performance and may especially follow different measuring principles \cite{prob_robotics_feat}, the interfaces must be very flexible, which the fusion and tracking algorithm has to cope with.
\item As the execution of driving maneuvers based on external data is a safety-critical task, high demands on the precision, reliability and the delay of the data are made. Thus, the system must deliver reliable and continuous tracks with the best available information and minimal latency.
\end{enumerate}

In conclusion to this, the key idea of this work comprises two parts, which follow from the principle limitations of real sensors. On the one hand, a generic interface is designed, which allows the transmission of incomplete measurements from sensors to the \ac{MEC} server, i.e. all measured quantities, as for example the extent or the speed of an object, are optional, except the position of the object. On the other hand, a multi-sensor multi-object filter is developed, which is not only capable of inferring the motion state of dynamic objects from the observation over time, as is done countless times by using a Kalman filter, but also capable of inferring the (optionally) missing information about the extent of the objects from measurements of multiple distributed sensors. Inference of this data is possible if sensors detect objects from different aspect angles and their position measurements refer to different points on the edge of an object. Within this work, the vertices of an object are used for this and called object reference points.


\subsection{Related Work} \label{related_work}

While the evolving mobile network standard \ac{5G} is not fully standardized and publicly available, infrastructure-based systems often utilize the ITS-G5 standard, enabling wireless ad-hoc networks between independent nodes without central control. As in \cite{lmb_t2t_1,lmb_t2t_2,lmb_t2t_3}\color{black}, the fusion and tracking mostly follows a mesh-like structure. Within these publications, sensor nodes with an own processing unit run the sensing, detection, multi-object association, fusion and tracking locally and then distribute this information to their neighbor nodes or a central node \cite{lmb_t2t_4}. These approaches lead to a simple interface and yet enable the utilization of highly sophisticated and sensor-specific tracking algorithms that allow the consideration of different sensor-specific characteristics. One example that specializes on radar and lidar is the \ac{RFS}-based extended multi-object target tracking approach of \cite{messmodell_scheel}.

The fusion of the resulting temporally filtered tracks from different sensors with unknown inter-sensor correlation is often done via \ac{GCI} \cite{GCI}, an elaborately investigated method \cite{GCI_interpretation,GCI_discussion}. However, \ac{GCI} solves the fusion problem only sub-optimally, since the calculation of the inter-sensor correlation between tracks is intractable. Moreover, the computationally expensive task of measurement-to-track and track-to-track association has to be calculated multiple times during local tracking and track-to-track fusion.

Since the transmission of raw data is, albeit the availability of large network bandwidths, still infeasible and often even impossible, \cite{ulm_generic_fusion} proposes a feature level approach. Advantageously, a central and therefore optimal fusion and tracking strategy can be followed \cite{Shalom_TF}, and the twofold execution of the computationally expensive measurement-to-track and track-to-track association within the sensors' processing units and the fusion unit can be avoided. The tracking is mostly done utilizing the multi-object extension of the Bayes filter \cite{fisst_fusion}, which models the multi-object states by an \ac{RFS}. Different filters have been developed, at which \ac{MHT} \cite{mht_tracking}, \ac{CPHD} \cite{cphd_tracking} and \ac{LMB} \cite{lmb_reuter} filters have become a quasi standard.

\subsection{Contributions}
Based on the requirements due to the chosen system structure, the rather unusual approach of a centralized fusion and tracking is treated within this work. A generic and flexible communication interface is proposed that utilizes an extended version of the feature level approach to respect the limited sensor capabilities. It allows to measure only a subset of the state space. Moreover, the proposed multi-sensor multi-object \ac{LMB} filter allows for the inference of these unmeasured information by utilization of several object measurements with different aspect angles.

The paper is structured as follows: The structural aspects and details about the considered interface are given in Section \ref{system_architecture}. Afterwards, details about the centralized multi-sensor multi-object fusion and tracking, utilizing a \ac{LMB} filter, are given in Section \ref{lmb}. The publication is completed by a both simulative and practical evaluation in Section \ref{evaluation} and a conclusion in Section \ref{conclusions}.

\section{Notation}
Within this work, individual objects are modeled as Gaussian distributions $\mathcal{N}(\underline{x};\hat{\underline{x}},\underline{P})$ under the assumption of a flat world, where probability densities are denoted by $p(\underline{x})$. Object detections, i.e. the measurements from the sensors, and tracks, i.e. objects that have been observed over a certain time by a filter, are distinguished. While an object of the former type may possibly be incomplete, i.e. only a subset of its features, as described in the following, is measured, tracks are always complete. Both types are described by a rectangular shaped object model, which is defined by the features $o = [\underline{x}, \underline{P}, \zeta, r, \ell, t, t', k]^T$. Here, $\underline{x} = [x_{\zeta}, y_{\zeta}, \varphi, v, w, l, h]^T$ describes the motion state of the object and $\underline{P}$ its uncertainty in form of the corresponding covariance matrix. The positions $x_{\zeta}$ and $y_{\zeta}$ are measured in relation to a specific object reference point $\zeta \in \mathbb{A}$, with $\mathbb{A} = \{\mathsf{FL}, \mathsf{FR}, \mathsf{BL}, \mathsf{BR}\}$ being the set of defined object reference points. Hence, sensors have to relate the position of an object to one of the vertices of the rectangular model, i.e. the front left or right or the back left or right. In exception to that, the position of tracks is related to the center of the rectangle. The orientation angle $\varphi$ and the object speed $\underline{v} = [v_{x}, v_{y}]^T$ always refer to the rectangle's center, too. Furthermore, $w$, $l$ and $h$ denote the width, length and height of the object, $\ell$ is a unique label and $k$ is a time step index. Respectively, $r = p(\exists x)$ denotes the probability of existence of the object, $t \in \mathbb{T}$ is its class, where $\mathbb{T}$ is the set of defined classes, and $t'$ is the probability of the object being of the given type $t$. For the sake of simplicity, $\mathbb{T}$ is, within this work, restricted to the classes \textit{car} and \textit{unknown}, but is arbitrarily extensible.

Furthermore, within this work the notation introduced by \cite{glmb_tracking} is used in the context of \ac{LMB} filters.


\section{SYSTEM ARCHITECTURE} \label{system_architecture}
In order to communicate an environment model and measurements, common interfaces are required. This section gives an overview about the structural aspects of the system and, afterwards, defines the environment model and the communication interface, which both reflect the requirements of generality and flexibility.

\subsection{Structural Aspects}
\begin{figure}[bp]
\centering
\begin{tikzpicture}
\node[rectangle, draw, thin, fill=white, double copy shadow, minimum height=0.5cm, text width=1.0cm, align=center](sensors) {\scriptsize Sensors};
\node[right=0.5cm of sensors, rectangle, draw, thin ,fill=white, double copy shadow, minimum height=0.5cm, text width=1.0cm, align=center](proc) {\scriptsize Processing};
\draw[-latex, double copy shadow] (sensors) -- (proc);
\node[right=0.5cm of sensors, rectangle, draw, thin ,fill=white, double copy shadow, minimum height=0.5cm, text width=1.0cm, align=center](proc) {\scriptsize Processing};

\node[yshift=2pt, right=1.0cm of proc, rectangle, draw, thin, fill=white, minimum height=0.5cm, text width=1.5cm, align=center](lmb) {\scriptsize LMB-Filter};
\draw[-latex, dashed] (proc.east) -- ([yshift=-2pt]lmb.west);
\draw[-latex, dashed] ([yshift=2pt, xshift=2pt]proc.east) -- ([yshift=0pt]lmb.west);
\draw[-latex, dashed] ([yshift=4pt, xshift=4pt]proc.east) -- ([yshift=2pt]lmb.west);

\node[right=1.0cm of lmb, rectangle, draw, thin, fill=white, minimum height=0.5cm, text width=1.0cm, align=center](vehicle) {\scriptsize Vehicle};
\draw[-latex, dashed] (lmb.east) -- (vehicle.west);

\node[rectangle](net1) at ($(proc)!0.5!(lmb)$) {};
\draw[dashed] ([yshift=17pt]net1.north) -- ([yshift=-12pt]net1.south);
\node[rectangle](net2) at ($(lmb)!0.5!(vehicle)$) {};
\draw[dashed] ([yshift=17pt]net2.north) -- ([yshift=-12pt]net2.south);

\node[anchor=north, inner sep=1pt] at ([yshift=24pt]$(sensors)!0.5!(proc)$) {\scriptsize Infrastructure Sensors};
\node[anchor=north, inner sep=1pt] at ([yshift=22pt]$([yshift=2pt]proc)!0.5!(vehicle)$) {\scriptsize MEC-Server};
\node[anchor=north, inner sep=1pt] at ([yshift=22pt]vehicle) {\scriptsize Receiver};

\node[inner sep=1pt, text width=1.2cm, align=center] at ([yshift=-25pt]$(proc)!0.5!(lmb)$) {\begin{spacing}{0.5}\scriptsize generic object list interface\end{spacing}};
\node[inner sep=1pt, text width=1cm, align=center] at ([yshift=-25pt]$(lmb)!0.5!(vehicle)$) {\begin{spacing}{0.5}\scriptsize track list interface\end{spacing}};

\end{tikzpicture}%
\caption{System overview with three parts, the distributed infrastructure sensors with associated processing units, MEC-Server, hosting the multi-sensor multi-object tracking unit (\ac{LMB} filter) and the receiver side, e.g. automated vehicles.}
\label{fig:system_overview}
\end{figure}
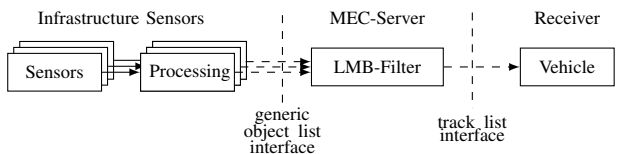

Cornerstone of the proposed system is a \ac{MEC} server, which is located at the edge of the local mobile network and offers very low communication delays. Furthermore, the system consists of distributed sensors with associated processing units, vehicles that receive the data, and a multi-sensor multi-object tracking unit (\ac{LMB} filter) that is implemented on the \ac{MEC} server. All components can be seen in Fig.~\ref{fig:system_overview}, where arrows indicate communication flow. Dashed arrows indicate a wireless connection and vertical dashed lines represent the used interfaces.

The principle flow of data is unilateral. Therefore, time-synchronized sensors (e.g. via \acl{GPS}, \acs{GPS}) transmit measurements to the \ac{MEC} server, which is, within this work, denoted as uplink. These measurements are collected, sorted into cycles of fixed time $T_{cycle}$, and then used to calculate an updated model of the environment every cycle. Finally, this model is transmitted to the vehicles, which is denoted as downlink. An inverse communication is not foreseen (ignoring technical communication, needed for control of connections).

The environment model of the system is confined to active and passive traffic participants, denoted as (dynamic) objects, and locally restricted. Hence, within a global context, the system has a cellular character with local zones, covered by independent systems that may communicate with each other, e.g. to pass over tracks using track-to-track fusion strategies. However, this is beyond the scope of this work and handled by aforementioned articles.

\subsection{Feature-level Interface}
The proposed feature-level interface defines the messages, which are communicated in the up- and downlink. In principle, these consist of two parts. First, a header, containing the timestamp (which is valid for the whole message) and a sensor ID, is sent, and, second, an informational part, which consists of lists of object detections (uplink) or tracks (downlink). Hereby, these messages follow the form $[\text{header}, \mathcal{O}]$, where $\mathcal{O} = {o_1, ... , o_n}$ is a list of object detections or tracks.

Furthermore, the definition of the interface differs a little between up- and downlink, since the two main requirements of allowing for maximal flexibility and enabling the inference of unmeasured state variables have to be respected. Based on the former, the interface has to support the simple connection of a wide range of sensors to the system. In order to achieve this, the description of a measurement in the uplink comprises of mostly optional features, i.e. only the position of an object has to measured, which supports a huge range of sensors. Contrary to object measurements, tracks are always described by the complete feature set. The demand for unique labels is relaxed in the uplink.

However, the second requirement cuts this flexibility slightly. It raises some soft constraints on the sensor setup and illumination, which are described in detail in the following subsection. Furthermore, regarding the interface in the uplink, position measurements must refer to any of the allowed object reference points, since the key idea of this work is to infer knowledge about the extent of an object due to measurements of different object sections from different, in the best case opposite, directions.

While this interface design has the drawback that, due to the missing feedback channel, some sensor-specific tracking schemes are impossible, state-of-the-art object detectors often create object proposals anyway \cite{nn_survey}. Moreover, the proposed interface allows for the use of various sensors from low-cost to high end, thus facilitates a widespread use of the system in a huge variety of venues.

In order to process measurements of the sensors, these are required to register and deregister at the \ac{MEC} server. Therefore, messages are available that contain the sensor's global position, type, orientation and a so-called covered area. The covered area is defined by a polygonal line $\mathcal{P}$ with the edge points $p_1, ..., p_m \in \mathbb{R}^2$, which are transmitted. The covered area defines the field of view of a sensor.

\subsection{Sensor Setup and Illumination}
There are only little requirements on the sensors. First, these must measure at least the positions $x$ and $y$ of objects, other features are optional, while each extra feature improves the robustness of the tracking. Second, the sensor must be aware of its global position, the common time base and its covered area.

However, the inference of unmeasured information makes more demands on the sensor setup. An area is sufficiently covered by sensors only if, in sum of the sensors' measurements, all features of the objects are measured or inferable. This, for example, would be the case if a single sensor measures the whole feature list with sufficient precision. However, taking into account that typical sensors measure only subsets of the feature vector \cite{prob_robotics_feat}, the requirements are relaxed, such that the total feature vector must be inferable by observations of the sensors altogether. This rather weak requirement leads to a set of rules that must be met:
\begin{enumerate}[i)]
	\item All areas should at least be covered by two sensors to sufficiently resolve occlusions and increase reliability due to redundancy.
    \item If sensors measure only a little number of object features, the number of independent sensors covering a certain area increases. Moreover, sensors must measure features at different object reference points and therefore most commonly need to be distributed in such way that an object is observed from different (in the best case opposite) aspect angles.
\end{enumerate}

Note that there is one more constraints about the sensors' capabilities. Although sensors are allowed to measure subsets of the full feature vector, the birth of objects requires the knowledge of all features and is therefore only possible if \begin{enumerate}[i)]
	\item a sensor measures the full feature vector $\underline{x}$, or
	\item a sensor measures the type of an object with high certainty. The state vector of the born object is then complemented with default values of the measured type.
\end{enumerate}
Consequently, all areas of the system must be observed by at least one sensor which satisfies either i) or ii).


\section{Generic centralized Multi-Sensor Multi-Object Fusion and Tracking} \label{lmb}
The \ac{FISST} \cite{fisst_fusion} provides a probabilistic framework to multi-sensor multi-object tracking applications. It extends the standard Bayes filter to the multi-object case utilizing \ac{RFS} to model the multi-object densities of objects. The posterior multi-object density is
\begin{equation}
\label{mot_update}
  \pi(X|Z) = \frac{g(Z|X) \pi(X)}{\int{g(Z|X)\pi(X)}\delta X} \quad ,
\end{equation}
where $g(Z|X)$ is the likelihood of the measurement $Z$ given the multi-object state $X$, where $X$ and $Z$ are sets. Since there is no closed-form solution for (\ref{mot_update}), the approximative \ac{LMB} filter \cite{lmb_reuter} is used, which combines computational efficiency with enhanced performance in dense multi-object situations and inherent track labeling and suits the needs of the scenario perfectly and is real-time capable even in large environments.

Following \cite{glmb_tracking}, the use of subscript time indices is neglected in order to increase readability, but, to avoid ambiguities, the index $k|k-1$ of predicted values is abbreviated using $+$ as subscript.

The single-object posterior state distribution is given by
\begin{equation} 
\label{single_object_posterior}
  p^{(\theta)}(\underline{x},\ell|Z) = \frac{p_+(\underline{x},\ell) \cdot \psi_Z(\underline{x},\ell;\theta)}{\eta_Z^{(\theta)}(\ell)} \quad ,
\end{equation}
where $\eta_Z^{(\theta)}(l)$ is a normalization factor and
\begin{equation}
\label{association_likelihood}
  \psi_Z(\underline{x},\ell;\theta) = \frac{p_D(\underline{x},\ell) \cdot g(\underline{z}_{\theta(\ell)}|\underline{x},\ell)}{\kappa(\underline{z}_{\theta(\ell)})}
\end{equation} 
denotes the generalized measurement likelihood for the association $\theta$, which associates a measurement $\underline{z}$ with a certain state $\underline{x}$. Furthermore, $p_+(\underline{x},\ell)$ is the predicted single-object state distribution, $p_D(\underline{x},\ell)$ the state dependent detection probability, and $\kappa(\underline{z}_{\theta(\ell)})$ the clutter intensity. Additionally,
\begin{equation}
  g(\underline{z}_{\theta(\ell)}|\underline{x},\ell) = \mathcal{N}(\underline{z}_{\theta(\ell)};\underline{h}(\underline{x}) \cdot \underline{x},\underline{R})
\end{equation}
denotes the likelihood of a measurement $z_{\theta(\ell)}$ given the object $\underline{x}$, where $\underline{h}(\underline{x}) \in \mathbb{R}^{d \times n}$ is the measurement matrix and $\underline{R} \in \mathbb{R}^{d \times d}$ the measurement covariance matrix.

In order to process - possibly incomplete - measurements from various sensors with different fields of view, three main adaptions have been made to the standard \ac{LMB} filter.

First, the different fields of view of the sensors are tackled by the detection probability $p_D(\underline{x})$. It is modeled state dependent and reflects if an object is within or outside of a sensor's covered area. Therefore, $p_D(\underline{x})$ is defined by
\begin{equation}
  p_D(\underline{x}) = 
  \begin{cases}
	\lambda_D & \quad \text{if } d(\underline{x}) < -r  \quad ,\\
	1-\lambda_D & \quad \text{if } d(\underline{x}) > r \quad ,\\
	-\frac{0.5 + \lambda_D}{r} d(\underline{x}) - 0.5 & \quad \text{otherwise} \quad .
  \end{cases}
\end{equation}
Thus, the detection probability is $\lambda_D$ if the object is within the covered area and $1-\lambda_D$ if not. The relaxation parameter $r$ defines a zone between both areas, where the detection probability decreases linearly from $\lambda_D$ within to $1-\lambda_D$ outside the covered area. Thereby, $d(\underline{x})$ denotes the signed minimal spatial distance between the polygonal line $\mathcal{P}$ and the object $\underline{x}$:

\begin{equation}
  d(\underline{x}) = \mathcal{W}(\underline{x},\mathcal{P}) \cdot \min(||\underline{x},\mathcal{P}||)
\end{equation}
with
\begin{equation}
\mathcal{W}(\underline{x},\mathcal{P}) = 
 \begin{cases} 
  1 & \quad \text{if }\underline{x}\text{ is outside }\mathcal{P} \quad ,\\ 
  -1 & \quad \text{if }\underline{x}\text{ is inside }\mathcal{P} \quad .
 \end{cases}
\end{equation}

Second, the number of clutter measurements is modeled as being Poisson distributed with $\lambda_{C}$, and the clutter measurements are distributed uniformly within the observable subset $\mathcal{F} \subset \mathbb{R}^d$ in the $d$-dimensional measurement space, with (hyper-)volume $\mathrm{vol}(\mathcal{F})$. Accordingly, the clutter intensity $\kappa(\underline{z}_C)$ is defined by
\begin{equation}
  \kappa(\underline{z}_C) = \lambda_{C} \cdot c(\underline{z}) = 
  \begin{cases}
	\frac{\lambda_{C}}{\mathrm{vol}(\mathcal{F})} & \quad \text{if } \underline{z} \in \mathcal{F} \quad ,\\
	0 & \quad \text{if } \underline{z} \notin \mathcal{F} \quad .
  \end{cases}
\end{equation}
Note that the observable subset introduces new parameters, e.g. minimal and maximal length or width of an object, which are modeled class dependent. However, these parameters have a simple technical meaning and are easier to choose than the intensity directly, which is an abstract parameter in the measurement space and dependent on the dimension $d$.

Third, since sensors are allowed to measure subsets of the state space, the \ac{LMB} filter must be able to infer the missing information. Here, especially the inference of the extent is of interest, since the quantities calculated by differentiation of the position are inferred by the Kalman filter anyway. The proposed approach bases on the idea that sensors refer the measurement of an object's position to the object reference point $\zeta$, which is seen the best by the sensor. If objects are detected from at least two different sensors with different object reference points, the corresponding extent can be inferred, since the information about the objects extent is encoded implicitly within both measurements.

Therefore, the measurement matrix $\underline{h}(\underline{x}) \in \mathbb{R}^{d \times n}$ must specify the transformation of an object's position from the center of the rectangular model, to which the position in the state space refers, to the respective object reference point $\zeta$. Here, without loss of generality, the state vector $\underline{x}$ is defined as $\underline{x} = [x, y, \ldots, w, l, h]^T \in \mathbb{R}^n$, i.e. the object's position is located at the state vector's beginning and the object's extent at the end, and arbitrary elements in between. $\underline{h}(\underline{x})$ is state dependent and non-linear.

In order to construct the measurement matrix, $\underline{\tilde{h}}(\underline{x}) \in \mathbb{R}^{n \times n}$ is defined first. Later, if $d < n$, i.e. a sensor measures a subset of the full state space only, the respective rows of $\underline{\tilde{h}}$ are to be deleted to obtain $\underline{h}$.

Since the measurement space and the state space only differ slightly, $\underline{\tilde{h}}$ is very similar to the identity matrix $\underline{I}$, and is defined as
\begin{equation}
  \underline{\tilde{h}}(\underline{x}) = 
  \begin{bmatrix}
    \underline{I} & \underline{\Delta}(\underline{x}) \\
    \underline{0} & \underline{I}
  \end{bmatrix} \quad .
\end{equation}
Furthermore, $\underline{\Delta}(\underline{x}) \in \mathbb{R}^{2 \times (n-2)}$ is responsible for the mentioned transformation of an object's position from the gravity center to the reference point $\zeta$. It is defined as
\begin{equation}
  \underline{\Delta}(\underline{x}) = 
  \begin{bmatrix}
    \underline{0} & \underline{f}(\zeta) & 0
  \end{bmatrix} \quad ,
\end{equation}
with
\begin{equation}
  f(\zeta) = \frac{1}{2} \cdot
  \begin{bmatrix}
  	-\sin{(\varphi)} & \cos{(\varphi)} \\
    \cos{(\varphi)} & \sin{(\varphi)}\\
  \end{bmatrix}  \circ
  \begin{bmatrix}
  	\delta & \gamma \\
    \delta & \gamma\\
  \end{bmatrix} \quad \in \mathbb{R}^{2 \times 2},
\end{equation}
where $\circ$ denotes the element-wise multiplication and
\begin{equation}
  \delta = 
  \begin{cases}
    1 & \text{if } \zeta = \mathsf{BL}, \mathsf{FL} \, ,\\
    -1 & \text{if } \zeta = \mathsf{BR}, \mathsf{FR} \, ,\\
  \end{cases} \,
  \gamma = 
  \begin{cases}
    1 & \text{if } \zeta = \mathsf{FL}, \mathsf{FR} \, ,\\
    -1 & \text{if } \zeta = \mathsf{BL}, \mathsf{BR}\, .\\
  \end{cases}
\end{equation}
Note that position and structure of $\underline{\Delta}(\underline{x})$ depend on the arrangement of the state vector $\underline{x}$.

Since the transmission of a certain reference point $\zeta$ is not mandatory within the proposed system, the \ac{LMB} filter must estimate the most probable reference points. Based on the viewing angle of a sensor on an object, the three closest corner points are chosen and handled in the innovation step of the \ac{LMB} filter. The one having the smallest Mahalanobis Distance \cite{mahalanobis} between predicted measurement $\underline{\hat{z}}_+$ and measurement $\underline{z}$ is then fed to the \ac{LMB} update.

Note that the use of a non-linear extension of the standard Kalman filter is required, since $\underline{h}(\underline{x})$ is non-linear. Within this work, the \ac{UKF} \cite{UKF_filter} is used, but using other non-linear extensions to the Kalman filter would also be possible. Further, arbitrary process models are allowed in the prediction, as long as the estimation of all features of the state $\underline{x}$ is possible.


\section{EVALUATION} \label{evaluation}
In order to measure the performance of the whole system, its components have been evaluated individually on the basis of simulated data first. Following that, a proof of concept demonstration at the test site is presented.

\begin{figure*}
~\\[5pt]
\centering
\input{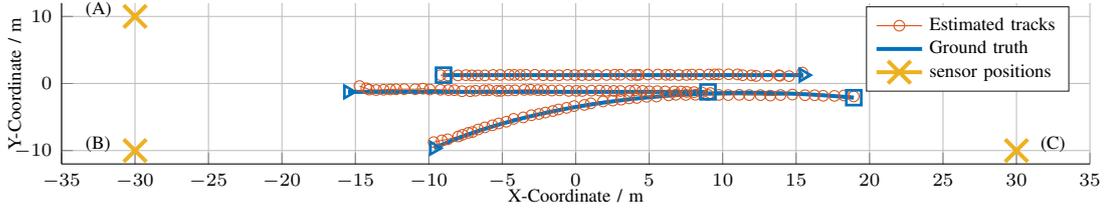}
\caption{Single simulation run of scenario 1 with three objects observed by three distributed sensors and $\underline{R} = [1, 0.5, 0.5]^T$.}
\end{figure*}

\subsection{Simulation}
The performance of the tracking has been investigated by a simulated scenario with three distributed sensors and three vehicles. The scenario follows the situation of a T-junction, where two vehicles pass each other in opposite driving directions on the major road. A third vehicle merges into the major road right in front of one of the vehicles. In order to make associations in the tracking algorithm ambiguous, the distances between the vehicles are extremely small. The ground truth positions can be seen in Fig.~\ref{fig:scenario1}, where the triangles mark the start of the vehicles' trajectories and the squares their end positions. All objects move following a \ac{CTRA} model. The estimated path of a single run is drawn in red.

Three distributed sensors are simulated, one in the upper (A) and one in the lower (B) left corner as well as one in the lower right (C) corner, as can also be seen in Fig.~\ref{fig:scenario1}. The vehicles are within the covered area of all sensors the whole time. The simulation was repeated twice with varying measured feature vectors. Within scenario 1, all sensors measure $\underline{z} = [x,y]^T$. In scenario 2, all sensors additionally measure either the objects' width or length, i.e. (A) and (C) measure the width, (B) the length. In both scenarios, the particular covariance matrix $\underline{R} = \text{diag}(\sigma_x, \sigma_y, \sigma_{w/l})$ is transmitted, too. Its elements fulfill $\sigma = \sigma_x = 2\sigma_y = 2\sigma_{w/l}$ and express the standard deviation in longitudinal and lateral direction in sensor coordinates as well as of the extent.

The measurements are created by adding zero-mean white Gaussian noise to the ground truth, whereat the standard deviation corresponds to the values in $\underline{R}$. A single measurement is removed by the chance $\nicefrac{1}{p_D}$ with $p_D = 0.95$, and uniformly distributed clutter measurements are added with rate $\lambda_C = 0.1$. The state vector in the \ac{LMB} filter is described by $x = [x,y,\varphi,\dot{\varphi},v,\dot{v},w,l]^T$, and a \ac{CTRA} process model is used in the \ac{LMB} filter.

Each simulation run was repeated three times, where the standard deviation of the additive white noise was increased from $\sigma = 0.5$ to $1.5$. Thus, six constellations are taken into account, which all have been repeated in 100 Monte Carlo trials. The estimation result was evaluated with the \ac{OSPAT} \cite{ospat} with \ac{OSPAT} order $p = 1$ and cut-off $c = 300$, where only the position is incorporated. The results of scenario 1 and 2 are given in Fig.~\ref{fig:ospat1} and \ref{fig:ospat2} respectively.

\begin{figure}[tbp]
%
%
\definecolor{mycolor1}{rgb}{0.00000,0.44700,0.74100}%
\definecolor{mycolor2}{rgb}{0.85000,0.32500,0.09800}%
\definecolor{mycolor3}{rgb}{0.92900,0.69400,0.12500}%
\begin{tikzpicture}

\begin{axis}[%
width=0.761\columnwidth,
height=0.4\columnwidth,
at={(0\columnwidth,0\columnwidth)},
scale only axis,
xmin=0,
xmax=50,
xlabel style={font=\color{white!15!black}},
xlabel={Time / k},
ymin=-1,
ymax=200,
ylabel style={font=\color{white!15!black}},
ylabel={OSPAT / m},
axis background/.style={fill=white},
axis x line*=bottom,
axis y line*=left,
xmajorgrids,
ymajorgrids,
legend style={legend cell align=left, align=left, draw=white!15!black},
ticklabel style={font={\scriptsize}},legend style={font={\scriptsize}, scale={0.4}, row sep={-1.5pt}},xlabel style={font={\scriptsize}, at={(axis description cs:0.5,-0.1)}},ylabel style={font={\scriptsize}, at={(axis description cs:-0.07,0.5)}}
]
\addplot [color=mycolor1]
  table[row sep=crcr]{%
1	300\\
2	67.296515031782\\
3	58.0431845080349\\
4	57.8813880055325\\
5	56.816886557694\\
6	56.544715287839\\
7	55.3108341266845\\
8	52.5670105973445\\
9	52.3123582245232\\
10	49.3308360500512\\
11	43.6516730158408\\
12	38.0433280367954\\
13	32.1241407350436\\
14	28.0150781250945\\
15	25.6418245606004\\
16	20.5563780823485\\
17	18.1947799912323\\
18	14.8553284170546\\
19	11.7928019286813\\
20	11.9637764729602\\
21	9.64636864143826\\
22	8.59218081352425\\
23	7.78235220783487\\
24	5.71237919274078\\
25	5.16975033005263\\
26	4.33784423541675\\
27	5.48914231710405\\
28	6.66528732436903\\
29	5.6737902255848\\
30	5.6827826715128\\
31	7.68044141850724\\
32	6.70260020771789\\
33	6.65961740782943\\
34	7.82145855109029\\
35	6.84962516382776\\
36	9.68028668656106\\
37	10.5028207222264\\
38	11.1188420926629\\
39	10.1946529188358\\
40	12.276349957401\\
41	12.4137458975266\\
42	12.5100060088385\\
43	13.531301297328\\
44	13.6286176537735\\
45	15.6699993450037\\
46	16.7408081375263\\
47	17.8470625664166\\
48	18.9212585386826\\
49	19.9273110306926\\
50	20.0167918237272\\
};
\addlegendentry{$\sigma\text{: 0.5m}$}

\addplot [color=mycolor2]
  table[row sep=crcr]{%
1	300\\
2	132.908213743055\\
3	110.153638214143\\
4	109.111822655858\\
5	108.133263998994\\
6	107.060592287142\\
7	106.072529929129\\
8	101.124303825368\\
9	99.172695182236\\
10	97.1885924234628\\
11	97.9947961234222\\
12	94.3768760984121\\
13	88.0232243222191\\
14	85.383185109833\\
15	77.357624671982\\
16	69.777066724512\\
17	65.6303203128754\\
18	60.385159728153\\
19	50.5655137749841\\
20	45.5939646903012\\
21	41.0300345322031\\
22	41.0965048150602\\
23	36.9614901276826\\
24	35.0473651161847\\
25	38.8443044615353\\
26	35.8775238552883\\
27	33.5660422784267\\
28	38.0853317336646\\
29	33.2226351152426\\
30	35.5057624044704\\
31	35.4301619613314\\
32	45.3479352732355\\
33	44.8933093182618\\
34	43.2653985037493\\
35	47.8908374774463\\
36	49.4986022636995\\
37	56.5776406253249\\
38	57.3446371126376\\
39	60.7315445838097\\
40	65.4949281361042\\
41	70.9573328839843\\
42	78.434748218072\\
43	82.2685839107488\\
44	84.3405408851426\\
45	85.3134794261223\\
46	88.0864612708741\\
47	93.7605460938755\\
48	99.1171546495266\\
49	99.7073215169941\\
50	103.771167953537\\
};
\addlegendentry{$\sigma\text{: 1m}$}

\addplot [color=mycolor3]
  table[row sep=crcr]{%
1	300\\
2	168.684516914719\\
3	122.214818479298\\
4	107.369759269884\\
5	103.31627420924\\
6	102.341992673976\\
7	102.812838503261\\
8	99.4284724878745\\
9	99.4661974744626\\
10	98.2734502487875\\
11	92.1347962194671\\
12	93.2079612160801\\
13	88.2650605739553\\
14	79.4018797643441\\
15	78.782677338279\\
16	70.529683514818\\
17	70.6993090627416\\
18	60.9523318450383\\
19	56.1938569770475\\
20	49.8569081999418\\
21	47.5658911824197\\
22	43.7021432299105\\
23	45.8606880103146\\
24	42.3448635727715\\
25	40.9266111394741\\
26	42.3180846244045\\
27	43.5484845834769\\
28	45.4920101137473\\
29	50.4345509853637\\
30	50.6988488385005\\
31	51.7820917310537\\
32	56.0042171742473\\
33	60.6322567208436\\
34	64.6389923214697\\
35	67.1781702482489\\
36	69.6929843290554\\
37	72.4564502414385\\
38	77.6499434877493\\
39	75.256009509027\\
40	79.5972664494604\\
41	87.5634580233184\\
42	89.1804023864489\\
43	95.626916767108\\
44	109.282892175756\\
45	115.944973186272\\
46	121.942349796272\\
47	128.140100496745\\
48	143.445368794277\\
49	142.892838256893\\
50	151.080324498131\\
};
\addlegendentry{$\sigma\text{: 1.5m}$}

\end{axis}
\end{tikzpicture}%
\caption{OSPAT errors over 100 Monte Carlo runs for varying measurement noise from $\sigma = 0.5$ to $\sigma = 1.5$ and three sensors with measured feature vector $\underline{z} = [x,y]^T$. \ac{OSPAT} order $p = 1$ and cut-off $c = 300$.}
\label{fig:ospat1}
\end{figure}
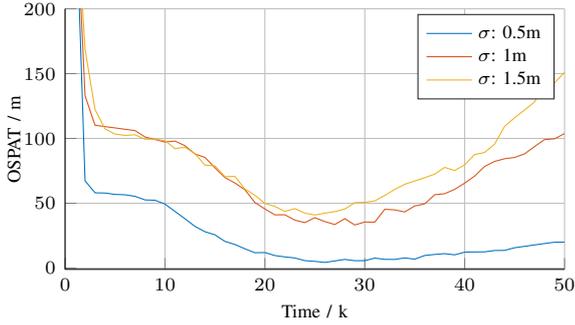
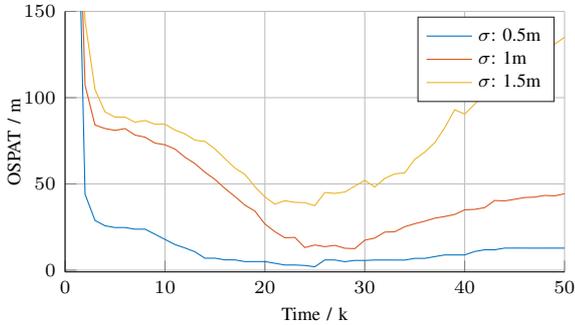
\begin{figure}[tbp]
%
%
\definecolor{mycolor1}{rgb}{0.00000,0.44700,0.74100}%
\definecolor{mycolor2}{rgb}{0.85000,0.32500,0.09800}%
\definecolor{mycolor3}{rgb}{0.92900,0.69400,0.12500}%
\begin{tikzpicture}

\begin{axis}[%
width=0.761\columnwidth,
height=0.4\columnwidth,
at={(0\columnwidth,0\columnwidth)},
scale only axis,
xmin=0,
xmax=50,
xlabel style={font=\color{white!15!black}},
xlabel={Time / k},
ymin=-1,
ymax=150,
ylabel style={font=\color{white!15!black}},
ylabel={OSPAT / m},
axis background/.style={fill=white},
axis x line*=bottom,
axis y line*=left,
xmajorgrids,
ymajorgrids,
legend style={legend cell align=left, align=left, draw=white!15!black},
ticklabel style={font={\scriptsize}},legend style={font={\scriptsize}, scale={0.4}, row sep={-1.5pt}},xlabel style={font={\scriptsize}, at={(axis description cs:0.5,-0.1)}},ylabel style={font={\scriptsize}, at={(axis description cs:-0.07,0.5)}}
]
\addplot [color=mycolor1]
  table[row sep=crcr]{%
1	300\\
2	44.0434797291706\\
3	28.8108611004496\\
4	25.7614837827145\\
5	24.7181664394175\\
6	24.7400484668494\\
7	23.762318393562\\
8	23.8002647476168\\
9	20.8304114415912\\
10	17.8479375781994\\
11	14.8823165327614\\
12	12.9003782228634\\
13	10.6680757993161\\
14	6.94894727354118\\
15	6.96647664378305\\
16	5.9964198508661\\
17	5.9831855399469\\
18	4.98842819430161\\
19	5.0025036140885\\
20	5.00236904563698\\
21	4.0080649107452\\
22	3.01178884001\\
23	3.01201667040655\\
24	2.74870545223329\\
25	1.98112474219425\\
26	5.94228036538014\\
27	5.92612171248632\\
28	4.93398354571253\\
29	5.65938224062576\\
30	5.6544313917259\\
31	5.9046982959572\\
32	5.9126098912601\\
33	5.89016393133917\\
34	5.86029932268585\\
35	6.85666324939435\\
36	6.85536871867127\\
37	7.87225733439852\\
38	8.85312710759234\\
39	8.84606146446391\\
40	8.84097888634671\\
41	10.8340748790221\\
42	11.8307573258571\\
43	11.8241735114423\\
44	12.8095840567658\\
45	12.8200510115706\\
46	12.8144318592755\\
47	12.7876447273187\\
48	12.7994562111242\\
49	12.7979155822739\\
50	12.7926878829091\\
};
\addlegendentry{$\sigma\text{: 0.5m}$}

\addplot [color=mycolor2]
  table[row sep=crcr]{%
1	300\\
2	107.194605633922\\
3	84.1333255808715\\
4	82.0957930988983\\
5	81.0482348890374\\
6	82.0651385223809\\
7	78.3196211067622\\
8	77.1056489542762\\
9	73.6667739161479\\
10	72.6768711511184\\
11	70.2029926600781\\
12	65.4988852233668\\
13	61.7583540300791\\
14	56.800808134004\\
15	52.5974915110499\\
16	47.3999630211581\\
17	42.6996675885618\\
18	37.7226399231588\\
19	34.2843450394131\\
20	26.8235679234772\\
21	22.3186006522991\\
22	18.8062083008465\\
23	19.020927086776\\
24	13.1234705971362\\
25	14.618549440708\\
26	13.6257436493983\\
27	14.4147281476296\\
28	12.7196647480469\\
29	12.4942555800763\\
30	17.4002004833074\\
31	18.6164746905078\\
32	22.0863646512784\\
33	22.2761377889368\\
34	25.2184562900129\\
35	26.9290453877561\\
36	28.4075086365605\\
37	30.157997834625\\
38	31.109647522384\\
39	32.3121297499952\\
40	35.0061371522257\\
41	35.2466868055677\\
42	36.2522880636173\\
43	40.4110993589485\\
44	40.1643980145125\\
45	41.0935611402995\\
46	42.1184750808335\\
47	42.3580437555701\\
48	43.3429340183197\\
49	43.0935533149899\\
50	44.2907263451455\\
};
\addlegendentry{$\sigma\text{: 1m}$}

\addplot [color=mycolor3]
  table[row sep=crcr]{%
1	300\\
2	144.596685599831\\
3	104.648045515939\\
4	91.7435456599876\\
5	88.7228625320362\\
6	88.7180956025285\\
7	85.7538087336277\\
8	86.7665365338835\\
9	84.60650692398\\
10	84.6337740724657\\
11	81.2101217162898\\
12	79.0197234638634\\
13	75.3363609219855\\
14	74.6416438756832\\
15	70.2749954255395\\
16	64.6330797326139\\
17	59.1978533291994\\
18	55.5189964875095\\
19	48.413628942814\\
20	42.4714708667978\\
21	38.3147727963991\\
22	40.2762232426595\\
23	39.330054791936\\
24	39.1250125213888\\
25	37.3954297324636\\
26	45.0173576598468\\
27	44.5258464575143\\
28	45.3111181917712\\
29	48.7431608800189\\
30	52.0499518200396\\
31	48.2513832463454\\
32	53.22039554711\\
33	55.8246360406377\\
34	56.4537880907185\\
35	64.2673494363608\\
36	68.3928806863388\\
37	73.9056943353873\\
38	82.5688665317504\\
39	92.9961168807761\\
40	90.4666214044072\\
41	96.2113522281934\\
42	102.244125531223\\
43	104.352054615821\\
44	113.000293368753\\
45	109.288807309507\\
46	109.71739924965\\
47	119.586625233266\\
48	125.794489477006\\
49	130.96852733368\\
50	135.009322343535\\
};
\addlegendentry{$\sigma\text{: 1.5m}$}

\end{axis}
\end{tikzpicture}%
\caption{OSPAT errors over 100 Monte Carlo runs and varying measurement noise from $\sigma = 0.5$ to $\sigma = 1.5$ and three sensors with measured feature vector $\underline{z}_{(A),(B)} = [x,y,w]^T$ and $\underline{z}_{(C)} = [x,y,l]^T$. \ac{OSPAT} order $p = 1$ and cut-off $c = 300$.}
\label{fig:ospat2}
\end{figure}
\begin{figure}[tbp]
%
%
\definecolor{mycolor1}{rgb}{0.00000,0.44700,0.74100}%
\definecolor{mycolor2}{rgb}{0.85000,0.32500,0.09800}%
\definecolor{mycolor3}{rgb}{0.92900,0.69400,0.12500}%
\begin{tikzpicture}

\begin{axis}[%
width=0.761\columnwidth,
height=0.4\columnwidth,
at={(0\columnwidth,0\columnwidth)},
scale only axis,
unbounded coords=jump,
xmin=0,
xmax=50,
xlabel style={font=\color{white!15!black}},
xlabel={Time / k},
ymin=3.5,
ymax=7.5,
ylabel style={font=\color{white!15!black}},
ylabel={Object length / m},
axis background/.style={fill=white},
legend style={legend cell align=left, align=left, draw=white!15!black},
ticklabel style={font={\scriptsize}},legend style={font={\scriptsize}, scale={0.4}, row sep={-1.5pt}},xlabel style={font={\scriptsize}, at={(axis description cs:0.5,-0.1)}},ylabel style={font={\scriptsize}, at={(axis description cs:-0.07,0.5)}}
]
\addplot [color=black, dashed]
  table[row sep=crcr]{%
0	0\\
};
\addlegendentry{Ground truth}

\addplot [color=black]
  table[row sep=crcr]{%
0	0\\
};
\addlegendentry{Estimated length}

\addplot [color=mycolor1, dashed, forget plot]
  table[row sep=crcr]{%
1	4.8\\
2	4.8\\
3	4.8\\
4	4.8\\
5	4.8\\
6	4.8\\
7	4.8\\
8	4.8\\
9	4.8\\
10	4.8\\
11	4.8\\
12	4.8\\
13	4.8\\
14	4.8\\
15	4.8\\
16	4.8\\
17	4.8\\
18	4.8\\
19	4.8\\
20	4.8\\
21	4.8\\
22	4.8\\
23	4.8\\
24	4.8\\
25	4.8\\
26	4.8\\
27	4.8\\
28	4.8\\
29	4.8\\
30	4.8\\
31	4.8\\
32	4.8\\
33	4.8\\
34	4.8\\
35	4.8\\
36	4.8\\
37	4.8\\
38	4.8\\
39	4.8\\
40	4.8\\
41	4.8\\
42	4.8\\
43	4.8\\
44	4.8\\
45	4.8\\
46	4.8\\
47	4.8\\
48	4.8\\
49	4.8\\
50	4.8\\
};
\addplot [color=mycolor1, forget plot]
  table[row sep=crcr]{%
1	nan\\
2	4.75\\
3	4.93\\
4	4.93\\
5	5.38\\
6	4.53\\
7	4.94\\
8	4.66\\
9	4.16\\
10	4.52\\
11	4.27\\
12	5.06\\
13	5.05\\
14	5.03\\
15	4.89\\
16	4.93\\
17	4.39\\
18	4.36\\
19	5.64\\
20	4.73\\
21	5.12\\
22	5.15\\
23	4.53\\
24	4.82\\
25	4.82\\
26	4.93\\
27	4.96\\
28	4.75\\
29	4.53\\
30	4.94\\
31	5.21\\
32	5.13\\
33	4.98\\
34	5.13\\
35	4.83\\
36	4.41\\
37	4.56\\
38	4.8\\
39	5.2\\
40	4.34\\
41	4.7\\
42	4.62\\
43	4.75\\
44	4.71\\
45	4.66\\
46	4.67\\
47	4.74\\
48	4.65\\
49	5.16\\
50	4.37\\
};
\addplot [color=mycolor2, dashed, forget plot]
  table[row sep=crcr]{%
1	5.2\\
2	5.2\\
3	5.2\\
4	5.2\\
5	5.2\\
6	5.2\\
7	5.2\\
8	5.2\\
9	5.2\\
10	5.2\\
11	5.2\\
12	5.2\\
13	5.2\\
14	5.2\\
15	5.2\\
16	5.2\\
17	5.2\\
18	5.2\\
19	5.2\\
20	5.2\\
21	5.2\\
22	5.2\\
23	5.2\\
24	5.2\\
25	5.2\\
26	5.2\\
27	5.2\\
28	5.2\\
29	5.2\\
30	5.2\\
31	5.2\\
32	5.2\\
33	5.2\\
34	5.2\\
35	5.2\\
36	5.2\\
37	5.2\\
38	5.2\\
39	5.2\\
40	5.2\\
41	5.2\\
42	5.2\\
43	5.2\\
44	5.2\\
45	5.2\\
46	5.2\\
47	5.2\\
48	5.2\\
49	5.2\\
50	5.2\\
};
\addplot [color=mycolor2, forget plot]
  table[row sep=crcr]{%
1	nan\\
2	4.99\\
3	5.35\\
4	5.47\\
5	5\\
6	5.02\\
7	5.63\\
8	5.45\\
9	5.01\\
10	5.32\\
11	5.97\\
12	5.73\\
13	5.05\\
14	5.06\\
15	5.42\\
16	5.55\\
17	5.53\\
18	5.19\\
19	5.38\\
20	5.14\\
21	5.01\\
22	5.39\\
23	5.33\\
24	4.88\\
25	4.83\\
26	5.53\\
27	4.91\\
28	5.73\\
29	5.46\\
30	5.51\\
31	5.04\\
32	5.26\\
33	5.05\\
34	5.26\\
35	5.19\\
36	5.11\\
37	5.05\\
38	5.4\\
39	5.33\\
40	4.96\\
41	4.9\\
42	5.1\\
43	4.89\\
44	5.11\\
45	5.35\\
46	5.32\\
47	5.2\\
48	5.09\\
49	5.24\\
50	5.69\\
};
\addplot [color=mycolor3, dashed, forget plot]
  table[row sep=crcr]{%
1	4.5\\
2	4.5\\
3	4.5\\
4	4.5\\
5	4.5\\
6	4.5\\
7	4.5\\
8	4.5\\
9	4.5\\
10	4.5\\
11	4.5\\
12	4.5\\
13	4.5\\
14	4.5\\
15	4.5\\
16	4.5\\
17	4.5\\
18	4.5\\
19	4.5\\
20	4.5\\
21	4.5\\
22	4.5\\
23	4.5\\
24	4.5\\
25	4.5\\
26	4.5\\
27	4.5\\
28	4.5\\
29	4.5\\
30	4.5\\
31	4.5\\
32	4.5\\
33	4.5\\
34	4.5\\
35	4.5\\
36	4.5\\
37	4.5\\
38	4.5\\
39	4.5\\
40	4.5\\
41	4.5\\
42	4.5\\
43	4.5\\
44	4.5\\
45	4.5\\
46	4.5\\
47	4.5\\
48	4.5\\
49	4.5\\
50	4.5\\
};
\addplot [color=mycolor3, forget plot]
  table[row sep=crcr]{%
1	nan\\
2	4.87\\
3	4.4\\
4	4.33\\
5	4.83\\
6	4.39\\
7	4.68\\
8	5.06\\
9	5.06\\
10	4.73\\
11	3.95\\
12	3.8\\
13	4.08\\
14	3.99\\
15	3.92\\
16	4.41\\
17	4.12\\
18	4.62\\
19	4.29\\
20	4.4\\
21	4.55\\
22	4.79\\
23	4.38\\
24	4.17\\
25	4.59\\
26	5.16\\
27	5.02\\
28	4.57\\
29	4.33\\
30	4.46\\
31	4.48\\
32	4.27\\
33	4.31\\
34	4.85\\
35	4.59\\
36	4.56\\
37	4.09\\
38	4.66\\
39	4.8\\
40	4.93\\
41	4.61\\
42	4.11\\
43	4.65\\
44	4.61\\
45	3.96\\
46	3.95\\
47	3.68\\
48	4.16\\
49	4.6\\
50	4.41\\
};
\end{axis}
\end{tikzpicture}%
\caption{Single simulation of scenario 1 with three objects observed by three distributed sensors and $\underline{R} = [1, 0.5, 0.5]^T$}
\label{fig:inference}
\end{figure}
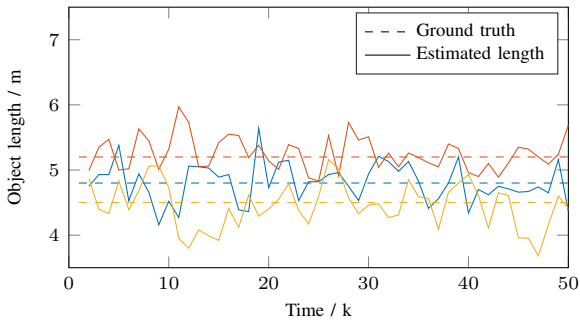

The results clearly show that the intended inference of information due to distributed sensors is possible. Moreover, it can be seen that the measurement of additional features increases the performance of the tracking regarding the reliability and stability against noise.

In the beginning of the \ac{OSPAT} curves from $k \approx 3$ to $10$, a shoulder-shaped region can be seen, where the \ac{OSPAT} error remains almost constant. Furthermore, at $k \approx 25$, an increment of the \ac{OSPAT} error can be seen. While the former error results from inaccurate track birth due to noise and indicates that the birth of tracks is crucial to the tracking, the second error arises after the vehicles pass each other and results from ambiguous measurement-to-track associations. However, both reveal that erroneous system states live on for a certain time until being corrected or released. This can be explained by noisy measurements, which lead to a 'wrong' reference point estimation in the \ac{LMB} filter, and therefore wrong estimation of the object's extent. If noise is strong or the measurement space is reduced, this situation occurs more often.

Thus, the limits of inference can be seen here, too. While the inference is possible with little noise, it meets its limits when standard deviation of the additional noise reaches half the smallest extent.

A more detailed look on one single run in Fig.~\ref{fig:inference} shows the progression of the estimated length of the objects in the case of correct measurement-to-track association. As can be seen, the inference is possible, but large jitter due to measurement noise is present. It shows that the inference of information is not smooth and is mainly driven by the measurement noise. The \ac{MSE} of the estimated object length in this example is within the order of magnitude of the measurement noise.

In order to tackle the problems in the case of heavy noise, further work should investigate the effect of using additional knowledge, for example a maximal yaw rate or a maximal length to width ratio of objects, within the measurement update.

\subsection{Real Data}
While the simulation shows that the inference of information from incomplete measurements is possible if all requirements are met, the practical example shall show that it is possible in reality, too. Therefore, the installation at the test site in Ulm-Lehr was used to create a proof-of-concept demonstration of the system. Related to the use case of resolving the occlusion, a scenario was created with a vehicle on the major road with the right of way heading towards an occluded T-junction. Three monocular cameras and two simple lidars that only emit 16 static beams are used \cite{digital_mirror}. The lidars are positioned at the particular end of the street and are directed towards each other. Both together cover most of the range, except the very left, where only one camera detects objects. The other cameras are distributed in the left and center part of the street and oriented in the driving direction of the vehicle.

\begin{figure}[tbp]
~\\[5pt]
\includegraphics[width=1.0\linewidth]{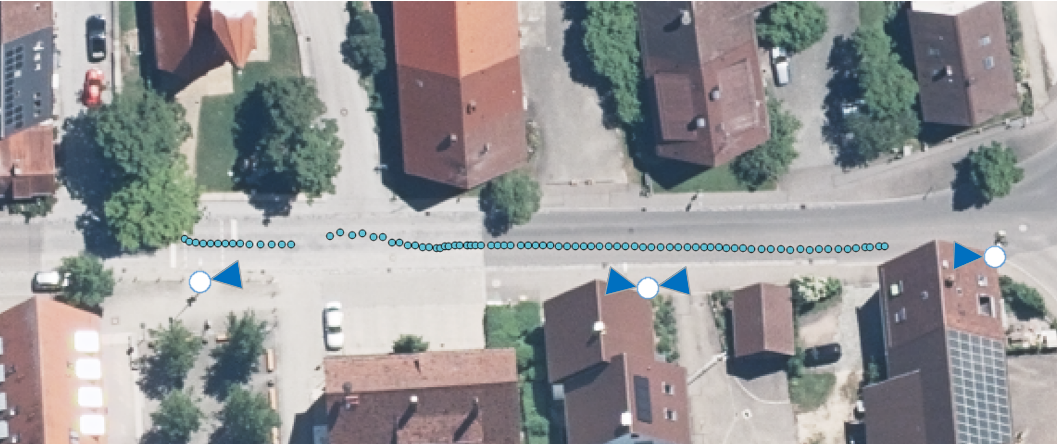}
\caption{Proof-of-concept demonstration of the environment modeling at a real test site in Ulm-Lehr. The track of a single vehicle is shown in cyan, the poles of the sensors are marked by white circles and blue cones, which indicate their field of view. \copyright ~Aerial photography: City of Ulm}
\label{fig:real_test}
\end{figure}

As can be seen in Fig.~\ref{fig:real_test}, the goal of following the vehicle's position could be solved. When the vehicle enter the system's observed area, the estimated position shows a tendency to the lane center, which is corrected after the vehicle is seen by multiple sensors.

Thus, a continuous track estimation between multiple sensor field of views is shown, where two different sensor types are used. Moreover, the sensors did measure only small subsets of the measurement space with different uncertainty. As the cameras delivered relatively precise lateral information but vague longitudinal information, the lidars behaved oppositionally. In the end, both sensor types complemented each other very well. However, since no ground truth is available, a more detailed examination is not possible, but it could be shown that the system is working on a real example. 

\section{CONCLUSIONS} \label{conclusions}
This paper introduced an interface for centralized feature-level based infrastructure systems for dynamic object detection. It is novel in its flexibility, since it allows to use sensors with different sensing capabilities far below the actual requirements of the state space. Such incomplete measurements can be processed by an adequately adapted \ac{LMB} filter. It uses the implicit information about the extent of objects from measurements of distributed sensors, if these measurements are related to different object reference points. The performance of the system has been shown in simulations and on a real world demonstration.

\addtolength{\textheight}{-10cm}   




\bibliographystyle{IEEEtran}
\bibliography{IEEEabrv,bibliography}

\begin{thebibliography}{10}
\providecommand{\url}[1]{#1}
\csname url@samestyle\endcsname
\providecommand{\newblock}{\relax}
\providecommand{\bibinfo}[2]{#2}
\providecommand{\BIBentrySTDinterwordspacing}{\spaceskip=0pt\relax}
\providecommand{\BIBentryALTinterwordstretchfactor}{4}
\providecommand{\BIBentryALTinterwordspacing}{\spaceskip=\fontdimen2\font plus
\BIBentryALTinterwordstretchfactor\fontdimen3\font minus
  \fontdimen4\font\relax}
\providecommand{\BIBforeignlanguage}[2]{{%
\expandafter\ifx\csname l@#1\endcsname\relax
\typeout{** WARNING: IEEEtran.bst: No hyphenation pattern has been}%
\typeout{** loaded for the language `#1'. Using the pattern for}%
\typeout{** the default language instead.}%
\else
\language=\csname l@#1\endcsname
\fi
#2}}
\providecommand{\BIBdecl}{\relax}
\BIBdecl

\bibitem{Co-PER}
M.~{Goldhammer}, E.~{Strigel}, D.~{Meissner}, U.~{Brunsmann}, K.~{Doll}, and
  K.~{Dietmayer}, ``Cooperative multi sensor network for traffic safety
  applications at intersections,'' in \emph{15th International IEEE Conference
  on Intelligent Transportation Systems}, 2012, pp. 1178--1183.

\bibitem{drive_C2X}
A.~Festag, L.~Le, and M.~Goleva, ``Field operational tests for cooperative
  systems: A tussle between research, standardization and deployment,'' in
  \emph{Proceedings of the Annual International Conference on Mobile Computing
  and Networking}, 2011, pp. 73--78.

\bibitem{simTD}
H.~{St{\"u}bing}, M.~{Bechler}, D.~{Heussner}, T.~{May}, I.~{Radusch},
  H.~{Rechner}, and P.~{Vogel}, ``simtd: a car-to-x system architecture for
  field operational tests [topics in automotive networking],'' \emph{IEEE
  Communications Magazine}, vol.~48, no.~5, pp. 148--154, 2010.

\bibitem{ETSI_MEC}
{ETSI}, ``Multi-access edge computing {(MEC)}; framework and reference
  architecture,'' {European Telecommunications Standards Institute (ETSI)},
  Sophia Antipolis Cedex, France, Group Specification (GS) MEC 002, 2019.

\bibitem{mecview}
\BIBentryALTinterwordspacing
R.~W. Henn. (2019) {MEC-View} project homepage. [Online]. Available:
  \url{http://www.mec-view.de/}
\BIBentrySTDinterwordspacing

\bibitem{ICT4CART}
\BIBentryALTinterwordspacing
A.~Amditis. (2019) {ICT4CART} project homepage. [Online]. Available:
  \url{https://www.ict4cart.eu/}
\BIBentrySTDinterwordspacing

\bibitem{digital_mirror}
M.~Buchholz, M.~Herrmann, J.~C. M{\"u}ller, V.~Belagiannis, P.~Pavlov,
  B.~H{\"a}tty, S.~Schulz, and R.~W. Henn, ``A digital mirror: A mobile edge
  computing service based on infrastructure sensors,'' presented at the 25th
  ITS World Congress Copenhagen, 2018.

\bibitem{prob_robotics_feat}
S.~Thrun, W.~Burgard, and D.~Fox, \emph{Probabilistic Robotics (Intelligent
  Robotics and Autonomous Agents)}.\hskip 1em plus 0.5em minus 0.4em\relax
  Cambridge, MA: The MIT Press, 2005, ch.~6.

\bibitem{lmb_t2t_1}
S.~{Li}, G.~{Battistelli}, L.~{Chisci}, W.~{Yi}, B.~{Wang}, and L.~{Kong},
  ``Multi-sensor multi-object tracking with different fields-of-view using the
  lmb filter,'' in \emph{21st International Conference on Information Fusion},
  2018, pp. 1201--1208.

\bibitem{lmb_t2t_2}
C.~{Fantacci}, B.~{Vo}, B.~{Vo}, G.~{Battistelli}, and L.~{Chisci}, ``Robust
  fusion for multisensor multiobject tracking,'' \emph{IEEE Signal Processing
  Letters}, vol.~25, no.~5, pp. 640--644, 2018.

\bibitem{lmb_t2t_3}
B.~{Wang}, W.~{Yi}, R.~{Hoseinnezhad}, S.~{Li}, L.~{Kong}, and X.~{Yang},
  ``Distributed fusion with multi-bernoulli filter based on generalized
  covariance intersection,'' \emph{IEEE Transactions on Signal Processing},
  vol.~65, no.~1, pp. 242--255, 2017.

\bibitem{lmb_t2t_4}
A.~K. {Gostar}, T.~{Rathnayake}, A.~{Bab-Hadiashar}, G.~{Battistelli},
  L.~{Chisci}, and R.~{Hoseinnezhad}, ``Centralized multiple-view information
  fusion for multi-object tracking using labeled multi-bernoulli filters,'' in
  \emph{2018 International Conference on Control, Automation and Information
  Sciences}, 2018, pp. 238--243.

\bibitem{messmodell_scheel}
A.~{Scheel}, S.~{Reuter}, and K.~{Dietmayer}, ``Vehicle tracking using extended
  object methods: An approach for fusing radar and laser,'' in \emph{2017 IEEE
  International Conference on Robotics and Automation}, 2017, pp. 231--238.

\bibitem{GCI}
M.~B. {Hurley}, ``An information theoretic justification for covariance
  intersection and its generalization,'' in \emph{Proceedings of the 5th
  International Conference on Information Fusion.}, vol.~1, 2002, pp. 505--511.

\bibitem{GCI_interpretation}
G.~Battistelli, ``An information-theoretic approach to distributed state
  estimation,'' 2011, pp. 12\,477--12\,482.

\bibitem{GCI_discussion}
T.~{Li}, J.~M. {Corchado}, and S.~{Sun}, ``On generalized covariance
  intersection for distributed {PHD} filtering and a simple but better
  alternative,'' in \emph{20th International Conference on Information Fusion},
  2017, pp. 1--8.

\bibitem{ulm_generic_fusion}
M.~{Munz}, M.~{Mahlisch}, and K.~{Dietmayer}, ``Generic centralized multi
  sensor data fusion based on probabilistic sensor and environment models for
  driver assistance systems,'' \emph{IEEE Intelligent Transportation Systems
  Magazine}, vol.~2, no.~1, pp. 6--17, 2010.

\bibitem{Shalom_TF}
Y.~Bar-Shalom, P.~K. Willet, and X.~Tian, \emph{Tracking and Data Fusion. A
  Handbook of Algorithms}, ser. Yakoov Bar-Shalom Publishing.\hskip 1em plus
  0.5em minus 0.4em\relax Storrs, CT: YBS Publishing, 2011.

\bibitem{fisst_fusion}
R.~P. Mahler, \emph{Statistical Multisource-Multitarget Information
  Fusion}.\hskip 1em plus 0.5em minus 0.4em\relax Norwood, MA, USA: Artech
  House, Inc., 2007.

\bibitem{mht_tracking}
D.~{Reid}, ``An algorithm for tracking multiple targets,'' \emph{IEEE
  Transactions on Automatic Control}, vol.~24, no.~6, pp. 843--854, 1979.

\bibitem{cphd_tracking}
R.~{Mahler}, ``Phd filters of higher order in target number,'' \emph{IEEE
  Transactions on Aerospace and Electronic Systems}, vol.~43, no.~4, pp.
  1523--1543, 2007.

\bibitem{lmb_reuter}
S.~{Reuter}, B.~{Vo}, B.~{Vo}, and K.~{Dietmayer}, ``The labeled
  multi-bernoulli filter,'' \emph{IEEE Transactions on Signal Processing},
  vol.~62, no.~12, pp. 3246--3260, 2014.

\bibitem{glmb_tracking}
B.~{Vo} and B.~{Vo}, ``Labeled random finite sets and multi-object conjugate
  priors,'' \emph{IEEE Transactions on Signal Processing}, vol.~61, no.~13, pp.
  3460--3475, 2013.

\bibitem{nn_survey}
\BIBentryALTinterwordspacing
D.~Feng, C.~Haase{-}Schuetz, L.~Rosenbaum, H.~Hertlein, F.~Duffhauss,
  C.~Glaser, W.~Wiesbeck, and K.~Dietmayer, ``Deep multi-modal object detection
  and semantic segmentation for autonomous driving: Datasets, methods, and
  challenges,'' \emph{CoRR}, vol. abs/1902.07830, 2019. [Online]. Available:
  \url{http://arxiv.org/abs/1902.07830}
\BIBentrySTDinterwordspacing

\bibitem{mahalanobis}
P.~C. {Mahalanobis}, ``On the generalized distance in statistics,''
  \emph{Proceedings of the National Institute of Sciences (Calcutta)}, vol.~2,
  pp. 49--55, 1936.

\bibitem{UKF_filter}
S.~J. Julier and J.~K. Uhlmann, ``New extension of the kalman filter to
  nonlinear systems,'' in \emph{Proceedings SPIE 3068, Signal Processing,
  Sensor Fusion, and Target Recognition VI}, 1997.

\bibitem{ospat}
B.~{Ristic}, B.~{Vo}, D.~{Clark}, and B.~{Vo}, ``A metric for performance
  evaluation of multi-target tracking algorithms,'' \emph{IEEE Transactions on
  Signal Processing}, vol.~59, no.~7, pp. 3452--3457, 2011.

\end{thebibliography}

\end{document}